\documentclass{article}
\usepackage{amsthm, graphicx}
\title{\Huge Desynched channels on IRCnet}
\author{Michael Hansen and Jeroen F. J. Laros\\
        \texttt{jlaros@liacs.nl}}
\date{\today}
\newtheorem{theorem}{Theorem}[subsection]
\theoremstyle{definition}
\newtheorem{example}[theorem]{Example}

\begin{document}

\maketitle

\begin{abstract} \noindent
In this paper we describe what a desynchronised channel on IRC is. We give
procedures on how to create such a channel and how to remove desynchronisation.
We explain which types of desynchronisation there are, what properties
desynchronised channels have, and which properties can be exploited.
\end{abstract}

\section{Introduction} \label{sec:intro}
IRC~\cite{IRC} is one of the oldest digital communication protocols on the
internet~\cite{IN}. This protocol is a form of synchronous conferencing which
is mainly used for its one-to-many communication capabilities. Although its
popularity has somewhat diminished since the introduction of instant messaging
applications \cite{IM} like the MSN messenger \cite{MSN}, it is still widely
used.

In this paper, we first give some background information about IRC in
Section~\ref{sec:bg}. In Section~\ref{sec:ds} we elaborate on the phenomenon of
\emph{desynched channels} and in Section~\ref{sec:place} we describe how to
place a boundary, Section~\ref{sec:properties} covers the occurrence of
\emph{fake modes} and Section~\ref{sec:back} describes how to remove a desync.
In Section~\ref{sec:app}, we give some examples of possible uses of fake modes
and we conclude in Section~\ref{sec:con}.

\section{Background} \label{sec:bg}
An IRC network consists of multiple servers connected to each other, there are
no cycles in this network, so the topology of this network is an undirected
tree (acyclic graph). An IRC network also has clients connected to its servers,
and messages are relayed from server to server to transfer a message from one
client to another.

This topology introduces some drawbacks; if a connection between two servers is
terminated, the graph will split into two parts. This is called a
\emph{netsplit}. When a server disconnects from the network, there are several
things that can happen; if the server in question is a leaf node, only the
server itself is split from the network. If on the other hand the server is
connected to $n$ other servers the network is split into $n + 1$ parts.

On IRC, there are so-called \emph{modes}, divided in ``user modes'' and
``channel modes'', which indicate rights and/or restrictions on users and
channels. A complete list of these modes can be found in \cite{IRC1, IRC2}.

A problem that can arise in a netsplit, is that a channel can have different
modes at each side of the split. This becomes a problem when the network is
rejoined and some action has to be taken to make sure the channel modes are
consistent throughout the network again.

Most problems concerning netsplits have been fixed already and we shall not
elaborate on this subject. We shall however investigate another less-known
drawback of the IRC topology known as \emph{desynching}.

We discuss this phenomenon in the context of IRCnet~\cite{IRN} only, many other
IRC networks do not have this problem, or to a lesser extent.

Since all messages (including mode changes) are propagated through the network,
there is no real concept of simultaneity. If for instance, at time $t_0$ client
$a$ sends a message and at time $t_1$ client $b$ sends a message, there is no
guarantee that a third client $c$ gets these messages in order. If client $c$
for example is physically closer to $b$, it might receive the message of $b$
before that of $a$ arrives. Normally these are only minor inconveniences but
things get worse if channel~mode changes are involved.

\section{Desynched channels} \label{sec:ds}
In general, a channel becomes \emph{desynched} if two incompatible mode changes
are propagated through the network at the same time.

There are two different cases to be distinguished.

\vbox{
\begin{enumerate}
\item Mode changes that are incompatible, but not mutually exclusive.
      \label{it:one}
\item Mode changes that are mutually exclusive. \label{it:two}
\end{enumerate}
}

\noindent
With mutual exclusiveness, we mean acting on each others mode or presence on
the channel.

\subsection{A ``flowing'' desync} \label{sec:flow}
\begin{figure}[h]
\includegraphics[width=\textwidth]{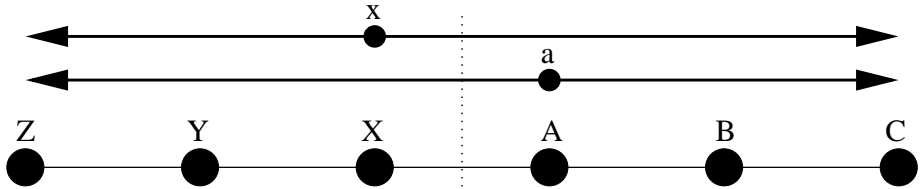}
\caption{A ``flowing'' desync} \label{fig:flow}
\end{figure}

\noindent
In case~\ref{it:one} the desynchronisation works as follows: Consider the
picture in Figure~\ref{fig:flow}, we have servers $A, B, C, X, Y$ and $Z$,
client $a$ on server $A$ and client $x$ on server $X$. Now both $a$ and $x$
issue an incompatible mode change at time $t_0$. Assume that it takes one time
unit for the change to propagate from one server to another.

\vbox{
\begin{enumerate}
\item At $t_0$ server $A$ will have accepted the mode change from $a$ and
      server $X$ that of $x$.
\item At $t_1$ server $B$ will have accepted the mode change from $a$ and
      server $Y$ that of $x$. At the same time, server $A$ accepts the mode
      change of $x$ and $X$ that of $a$, overwriting the previously set mode.
\item This process continues until $t_3$. At this time servers $X, Y$ and $Z$
      have accepted the mode $a$ has set and $A, B$ and $C$ have accepted the
      mode $x$ has set.
\end{enumerate}
}

\noindent
Notice that clients on this channel will see two mode changes following each
other, but not in the same order. Clients on the left side of the boundary (the
dotted line between server $X$ and $A$) will see the mode change of $x$ first,
then that of $a$. The clients on the other side will see the mode change of $a$
first, then that of $b$.

The channel is now desynched and the boundary is between server $X$ and $A$.

\vbox{
\begin{example}[A ``flowing'' desync in practice] \label{ex:flow}
We need two clients on different servers, like we have in
Figure~\ref{fig:flow}. Lets assume both have operator status on the channel in
question. Now at the same time, let them change the topic. Here is what client
$a$ will see:
\begin{verbatim}
-X- Topic (#channel): changed by a: I am a!
-X- Topic (#channel): changed by x: I am x!
\end{verbatim}

\noindent
And this is what client $x$ sees:
\begin{verbatim}
-X- Topic (#channel): changed by x: I am x!
-X- Topic (#channel): changed by a: I am a!
\end{verbatim}

\noindent
So now the topic looks different from the perspective of the servers $A, B$ and
$C$ on one side and $X, Y$ and $Z$ on the other side.
\end{example}
}

\noindent
The cases for which this mechanism applies are:

\vbox{
\begin{itemize}
\item \verb#key (+k)#.
\item \verb#limit (+l)#.
\item The \verb#private (+p)# / \verb#secret (+s)# combination.
\item \verb#topic#.
\item \verb#voice (+v)#.
\end{itemize}
}

\noindent
The following channel modes also ``flow'' in this way, except the server does
not propagate the mode if the mode change does not have any effect; the command
is rejected by the server. Therefore, a simultaneous mode change like this will
always result in the toggling of this mode.

\vbox{
\begin{itemize}
\item \verb#invite-only (+i)#.
\item \verb#moderated (+m)#.
\item \verb#no-external-messages (+n)#.
\item \verb#topic-control (+t)#.
\end{itemize}
}

\noindent
As a concluding remark, this type of desynching is relatively harmless, since
it is easy to get back in sync again and this type of desynching is
non-transferable. In Section~\ref{sec:back} we discuss how to get a desynched
channel back in sync and in Section~\ref{sec:col} we discuss the more dangerous
``colliding'' desynchronisation and the concept of transferable
desynchronisation.

\subsection{A ``colliding'' desync} \label{sec:col}
\begin{figure}[h]
\includegraphics[width=\textwidth]{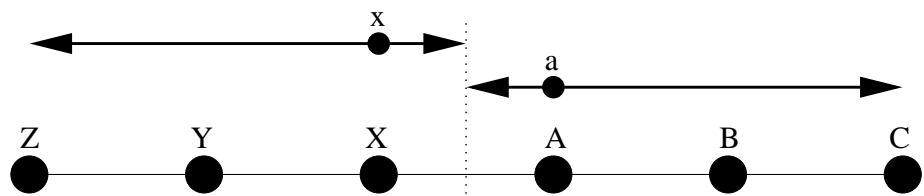}
\caption{A ``colliding'' desync} \label{fig:collide}
\end{figure}

\noindent
In case~\ref{it:two} the desynchronisation goes as follows: Consider the
picture in Figure~\ref{fig:collide}, again we have servers $A, B, C, X, Y$ and
$Z$, client $a$ on server $A$ and client $x$ on server $X$. Now both $a$ and
$x$ issue an incompatible and mutually exclusive mode change at the same time
$t_0$. Again, assume that it takes one time unit for the change to propagate
from one server to another.

\vbox{
\begin{enumerate}
\item At $t_0$ server $A$ will have accepted the mode change from $a$ and
      server $X$ that of $x$.
\item At $t_1$ server $B$ will have accepted the mode change from $a$ and
      server $Y$ that of $x$. The mode change of $a$ however does not
      propagate to $X$, neither does the one from $x$ propagate to $A$.
\item This process continues until $t_2$. At this time servers $X, Y$ and $Z$
      have accepted the mode $x$ has set and $A, B$ and $C$ have accepted the
      mode $a$ has set.
\end{enumerate}
}

\noindent
Notice that in this case the result is the opposite of what happens with a
``flowing'' desync. The clients on the channel in question only see one mode
change, the one on their side of the boundary.

\vbox{
\begin{example}[A ``colliding'' desync in practice] \label{ex:col}
Again, we need two clients on different servers, like we have in
Figure~\ref{fig:collide}. Lets assume both have operator status on the channel
in question. Now at the same time, let them deop each other. Here is what
client $a$ will see:
\begin{verbatim}
-+- mode/#channel [-o x] by a
\end{verbatim}

\noindent
And this is what client $x$ sees:
\begin{verbatim}
-+- mode/#channel [-o a] by x
\end{verbatim}

\noindent
So now $a$ has ops on the right side of the boundary, and $x$ has ops on the
left side.
\end{example}
}

\noindent
The cases for which this mechanism applies are:

\vbox{
\begin{itemize}
\item \verb#kick#.
\item \verb#ops (+o)# and by extension \verb#creator (+O)# (where applicable).
\end{itemize}
}

\noindent
Both of these cases result in having ops on one side of the boundary and not on
the other. At first glance, this only seems a disadvantage, but in
Section~\ref{sec:app} we show some possible uses.

\subsection{Other possible causes} \label{sec:othercause}
A netjoin can also cause desyncronisation of a channel. If for example, during
a netsplit a conflicting mode is set, a netjoin will result in a ``flowing''
desynchronisation of the channel.

Only ``flowing'' desyncs can occur on a netjoin and there are only a couple of
modes that can generate such a desynchronisation:

\vbox{
\begin{itemize}
\item \verb#key (+k)#.
\item \verb#limit (+l)#.
\item \verb#private (+p)# / \verb#secret (+s)#.
\item \verb#topic#.
\end{itemize}
}

\noindent
Notice that a conflict is not the absence of a mode versus the presence of one,
so a conflicting \verb#+l# mode is not \verb#+l# versus \verb#-l#, but for
example \verb#+l 10# versus \verb#+l 11#.

Also note that the \verb#+s# / \verb#+p# conflict is fixed in all operational
IRCnet servers.

\section{Placing the desync boundary} \label{sec:place}
Now we shall discuss how to place the desync boundary. To desync a channel, we
need two clients, $a$ and $b$, both on different servers. First, observe that
the desync boundary will always be between the servers $a$ and $b$ are on. The
easiest way to desync is by having two clients on adjacent servers. This way
the boundary will always be between those two servers. Using
Example~\ref{ex:col} will result in a desync.

\begin{figure}[h]
\includegraphics[width=\textwidth]{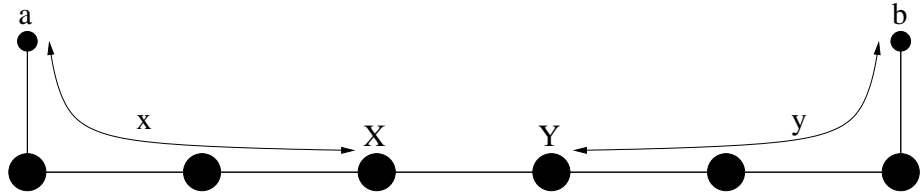}
\caption{Placing the desync boundary} \label{fig:place}
\end{figure}

\noindent
If $a$ and $b$ are not on adjacent servers, the boundary can be created
anywhere (as long as it is between the servers $a$ and $b$ are on). Where the
boundary will be created exactly is a matter of timing. By adjusting the delay
between the commands $a$ and $b$ give that will result in a desync, the
position of the boundary can be manipulated.

\vbox{
\vspace*{5mm}
\hrule
\vspace*{-1mm}
\begin{tabbing}
XXXXXXXXXXXXXXX\=XXX\=Xy\=Xy\=Xy\=Xy\=XXXXXXXX\=XXXXX\=XX \kill
  \>$\mathit{OneUserDesync}\,(X, Y, a, b) :: $\\
  \>\>$x := \mathit{MeasureLatency}\,(a, X);$\\
  \>\>$y := \mathit{MeasureLatency}\,(b, Y);$\\
  \>\>$\mathbf{if\ } x \le y \mathbf{\ then}$\\
  \>\>\>$\mathit{Command}\,(a);$\\
  \>\>\>$\mathit{Delay}\,(y - x);$\\
  \>\>\>$\mathit{Command}\,(b);$\\
  \>\>$\mathbf{else}$\\
  \>\>\>$\mathit{Command}\,(b);$\\
  \>\>\>$\mathit{Delay}\,(x - y);$\\
  \>\>\>$\mathit{Command}\,(a);$
\end{tabbing}
\vspace*{-1mm}
\hrule
\vspace*{1mm}
\centerline{Desynching with clients $a$ and $b$ under control of one user.}
\vspace*{1.2mm}
\hrule
\vspace*{5mm}
}

\noindent
Here we see a procedure that will desync a channel. This approach assumes that
clients $a$ and $b$ are controlled by one user.

First, the user chooses two adjacent servers on the path from $a$ to $b$. The
server closest to $a$ we call $X$ and the other one we call $Y$. Then client
$a$ measures the latency to server $X$, this can be done with the \verb#/ping#
command for example. Client $b$ does the same for server $Y$. Now, depending on
the difference in latency, client $a$ is kicked or deopped and after the
appropriate delay, client $b$ is kicked or deopped or vice versa.

\vbox{
\vspace*{5mm}
\hrule
\vspace*{-1mm}
\begin{tabbing}
XXXXXXXXXXXXXXX\=XXX\=Xy\=Xy\=Xy\=Xy\=XXXXXXXX\=XXXXX\=XX \kill
  \>$\mathit{TwoUserDesync}\,(X, a, b) :: $\\
  \>\>$x := \mathit{MeasureLatency}\,(a, X);$\\
  \>\>$y := \mathit{ReceiveLatency}\,(b);$\\
  \>\>$\mathit{SendLatency}\,(b, x);$\\
  \>\>$t := \mathit{AgreeOnTime}\,(b);$\\
  \>\>$\mathit{Delay}\,(t - \mathit{GetTime}\,())$\\
  \>\>$\mathbf{if\ } x \le y \mathbf{\ then}$\\
  \>\>\>$\mathit{Delay}\,(y - x);$\\
  \>\>$\mathit{Command}\,(b);$
\end{tabbing}
\vspace*{-1mm}
\hrule
\vspace*{1mm}
\centerline{Desynching with clients $a$ and $b$ under control of cooperating
            users.}
\vspace*{1.2mm}
\hrule
\vspace*{5mm}
}

\noindent
If the user does not have control over both clients, then two users must work
together to create a desync boundary. Both users must agree on the present time
however, so something like running NTP~\cite{NTP} will be needed.

First, both users must agree on two adjacent servers on the path to each other.
Each user now takes the server closest to it and calls it $X$. Then it measures
the latency to $X$ and communicates this to the other user. After this, they
agree on a time in the near future $t$ (of course, $t$ must be later than the
delay added to the present time). Now, depending on the difference in delay, on
time $t$ either client $a$ or $b$ will delay for the appropriate time before
issuing the kick or deop.

\section{Properties of a desynched channel} \label{sec:properties}
\subsection{Fake modes} \label{sec:fm}
Setting a mode on one side of a desynchronised channel will result in a
\emph{fake mode} on the other side. Having a fake voice (voice on one side, but
not on the other) is not very harmful, it is often even irritating to have one,
especially on a moderated channel. Only part of the other clients on the
channel will be able to see what the client with a fake voice has to say.
Furthermore, the fake voice client gets a ``cannot send to channel'' message
from the opposite side (the side it is not voiced on) each time it says
something. If the voice is on the opposite side and the channel is moderated, a
fake voice is useless.

Having a fake ops is more useful, because with this you can make other clients
a fake (op, voice) too, also channel modes can be faked (even if these channel
modes can not be used to desync themselves). A fake op can desynchronise the
topic for example, or the \verb#+m# setting. Every channel mode, including all
the lists (ban (\verb#+b#), ban-exception (\verb#+e#), etc) can be
desynchronised in this way. Leaving the channel in a very messy state.

\subsection{Fake joins} \label{sec:fj}
A fake join is an artifact of a kick on a desynchronised channel. It occurs
when we have the following situation:
\begin{center}
\verb#A-B|C#
\end{center}
Where the \verb#-# denotes a normal connection between servers and the \verb#|#
denotes the desync boundary. We have client $a$ on server $A$, client $b$ on
server $B$ and client $c$ on server $C$. Assume that client $b$ only has ops on
its side of the boundary. If client $b$ now kicks $a$, client $c$ will not see
$a$ leave (because from its perspective, $b$ does not have ops and therefore no
right to kick $a$). If $a$ now rejoins the channel, only $b$ will see him join
and since $c$ already sees him as joined, no new join is seen. Instead a ``fake
join'' error message is generated.

\subsection{The \tt{\&channel}} \label{sec:chan}
When a fake join or mode is generated, the error that results from it will be
logged on \verb#&channel#. This is one of the channels on the local server
where server messages are logged.

This error will be visible on all servers on the side the fake join was
generated. The servers at the other side see nothing but normal operation.

The error message itself includes the channel where the fake mode occurred and
the client that generated the fake mode.

The \verb#&channel# is also used for Reop notices.

\subsection{Determining where the boundary is} \label{sec:det}
There are a few tests to see where the boundary is. One is by giving a client
voice on one side of the boundary and setting channel mode \verb#+m# on the
other. When this client now sends a message, it will receive the error message
``cannot send to channel'' from some foreign server. The boundary lies between
that server and the next one, as seen from the perspective of the client.

Other possibilities are kicking a client from one side of the boundary and
setting channel mode \verb#+n# on the other. There are several more cases like
this, sometimes resulting in different error messages (like ``you're not on
that channel'').

\section{Getting a channel back in sync} \label{sec:back}
There are several ways to accomplish this, depending on the circumstances.
Synchronizing all channel modes and joined clients will resync the channel. In
most cases, a channel operator can repair the desynchronisation, provided there
are operators left on both sides of the boundary. If there are no channel
operators left, recreating the channel is one of the last options left. If
there are clients that have been kicked from one side of the boundary,
rejoining the channel will fix this.

Another, but less reliable way to repair a desynched channel is to wait for a
netsplit (and the consecutive netjoin). In some cases, this can get a channel
back in sync, or at least give a desynched ops control over the whole channel
again, making it able to repair the channel manually. This also applies to
semi-kicked clients, they will rejoin the channel as in a normal netjoin and
gain the privileges they had on their own side of the split.

\vbox{
\begin{example}[A netjoin repairing a desynched channel] \label{ex:netjoinrep}
Assume that we have the following situation:
\begin{center}
\verb#A-B|C#
\end{center}
Suppose we have client $a$ on server $A$ which has operator status on the
\verb#A-B# part of the channel and not on the other part.

If a netsplit occurs between $A$ and $B$ or between $B$ and $C$, a netjoin will
result in a join of client $a$ and giving him operator status because the
server will give him that.
\end{example}
}

\section{Applications} \label{sec:app}
We now give a few examples on what fake modes can be used for. We are only
covering a small subset of all possibilities and even these can be combined.

Also remember that in all cases, the actual act of desynching requires operator
status. The desynching itself might be done almost unseen, or it might take
several attempts, arousing suspicion.

\subsection{Preparing a takeover} \label{sec:prepare}
As mentioned in Section~\ref{sec:back}, a netjoin can resynchronise a channel.
This can be used to regain ops on the other side of the boundary. Here is what
can happen:

\vbox{
\begin{example}[Hiding ops] \label{ex:prepare}
Suppose a channel is desynched as follows:
\begin{center}
\verb#A|B-C#
\end{center}
Client $a$ resides on server $A$ and has lost ops on the rest of the channel.
This user can now bring in other clients on his side of the boundary and give
them ops on its side. The clients on the servers $B$ and $C$ see the clients
joining, but have no knowledge of modes being set (unless they monitor the
\verb#&channel#).

Now if a netsplit occurs between servers $A$ and $B$, on the netjoin the
clients that had ops on server $A$ will gain ops on the whole channel, since
the joining server gives it. This is a variation of the old splitsurfing (or
splitriding) idea.

This can even be taken one step further: if there is another client $b$
residing on server $B$ or $C$, which has lost its ops on server $A$, then $b$
can kick all the clients on server $A$ from its side of the channel, leaving
the ``kicked'' clients present only on server $A$. This will create a hidden
assembly of desynched ops, waiting for a netsplit to happen.
\end{example}
}

\noindent
User $a$ can even take further actions to increase the chance of a takeover. By
de-opping everyone at its side of the border, the operators on the other side
will have fake modes too. If these operators kick the clients $a$ brings in, it
looks to them like the channel is cleaned up. In reality it has no effect on
the side $a$ is on.

\subsection{Cloaking a client on a channel} \label{sec:cloak} As mentioned in
Example~\ref{ex:prepare}, a client can be made invisible by kicking them from
one side of the boundary. This can be used as an advantage in several ways.

\vbox{
\begin{example}[A cloaked client] \label{ex:cloak}
Suppose we have the following situation:
\begin{center}
\verb#A|B|C#
\end{center}
User $b$ is alone on server $B$ and is kicked from servers $A$ and $C$. This
client is now invisible for the clients residing on $A$ and $C$. If the channel
is not set to \verb#+n#, then $b$ is able to talk normally. However, $b$ will
only receive messages from the channel if there are clients on both servers $A$
and $B$, because the servers will not propagate messages if they think there
are no clients to propagate them to.

Modes however, are propagated to every server in the network. So if server $A$
is empty, client $b$ will only see the mode changes on the channel.
\end{example}
}

\noindent
In practice this can be used in large networks that have users in a few
countries, but not in the one of user $b$.

There is even the possibility to do this beforehand, by desynchronising the
channel, and then cleaning up the channel except for the cloaked client. When a
new client $a$ now joins on $A$ or $C$, this client will think it has created a
new channel and it gets a serverops as usual. This new client is now operator
on the whole channel, but does not know that $b$ is on there too. User $b$
might even have ops left or have a whole assembly of clients joined to server
$B$.

\subsection{Monitoring an anonymous channel} \label{sec:snoop}
On IRCnet, an uncollidable channel is one that has the \verb#!#-prefix
(\verb#!channel# for example). On such a channel anonymous mode (\verb#+a#) can
be set by the channel creator (who has mode \verb#+O#). When a channel has mode
\verb#+a# , everyone will have the nick \verb#anonymous# and the \verb#/names#
command will show only the nick of the one who gave the command. The \verb#+a#
mode can also be desynched, like any other. This can lead to the following
situation:

\vbox{
\begin{example}[A desynched anonymous channel] \label{ex:anon}
Below, we see a desynchronised channel.
\begin{center}
\verb#A|B#
\end{center}
User $a$ resides on server $A$ and has ops on server $A$ only. User $b$ is the
channel creator (\verb#+O#) and resides on server $B$. If $a$ now takes away
ops from $b$, then $b$'s operator status, as well as its creator status will
only be present on server $B$. If client $b$ now sets mode \verb#+a#, the
channel will be anonymous on $b$'s side, but not on $a$'s side. This results in
clients on $A$ seeing the real nicknames, while others won't.
\end{example}
}

\noindent
Notice that this two-server setup is only a simplified version, this idea is
more effective when instead of only one server ($B$) there rest of the IRC
network is on that side.

Also notice that client $a$ can set its part of the channel to invite-only
(\verb#+i#), thus preventing anyone joining the non-anonymous part of the
channel and finding out what is going on.

\subsection{Multiple desyncs} \label{sec:multi}
If a desynchronised channel is used for monitoring or cloaking, having multiple
boundaries could be useful. Consider the following:
\begin{center}
\verb#A|B|C#
\end{center}
Assume that client $a$ is on server $A$, and is cloaked (as described in
Section~\ref{sec:cloak}) from servers $B$ and $C$. Assume that server $B$ is
empty.

If the desync boundary between $A$ and $B$ now disappears (because of an
accident caused by client $a$, or a netsplit for example), it can be recreated
without people on server $C$ ever noticing that client $a$ was present. If the
boundary is recreated by clients that only have ops on servers $A$ and $B$, the
creation of the boundary will even be undetectable for people on server $C$.

Notice that if the boundary between servers $B$ and $C$ disappears, client $a$
will stay cloaked too, but recreating the boundary will be more difficult
because it will not go unnoticed for the clients residing on server $C$.

\section{Conclusion} \label{sec:con}
Tests have shown that the process of desynching a channel is fairly easy. Even
without using the procedures described in Section~\ref{sec:place}, but by using
a script where only the order of the commands were adjusted manually,
desynching was usually accomplished within two or three attempts. If the
boundary is not in a desired position, it can always be removed as described in
Section~\ref{sec:back} as long as there is a spare ops on the channel to get
the channel back in sync.

The position of the boundary can be manipulated to some extent, as explained in
Section~\ref{sec:place}. Determining where the boundary is, can be done as
described in Section~\ref{sec:det}, so when a boundary is not in the desired
place, it can be removed and another attempt can be made.

Having a desynched channel can have some advantages, as seen in
Section~\ref{sec:app}. It can, for example, be used to appear invisible on a
channel or for monitoring anonymous channels.

\section{Further research} \label{sec:further}
As mentioned in Section~\ref{sec:bg}, other IRC networks have not been
investigated thoroughly. We can report that on an Unreal~\cite{UNR} network
like Undermind~\cite{UND}, the desyncs that occur are of no use. The desync by
simultaneous deop seems to trigger a response from the servers that result in
both clients having ops on the wrong side of the desync boundary, rendering
them useless. Simultaneous kicks do not result in a desync at all, both clients
just get kicked. Other networks might or might not have desynchronisation
behaviour as described in this paper.

\section{Acknowledgements} \label{sec:ack}
The authors would like to thank the users on the IRCnet \verb#!ircd# channel
for their insights concerning fake joins.


\begin{thebibliography}{XX}
\bibitem{IRC} \verb#http://en.wikipedia.org/wiki/Internet_Relay_Chat#
\bibitem{IN} \verb#http://en.wikipedia.org/wiki/Internet#
\bibitem{IM} \verb#http://en.wikipedia.org/wiki/Instant_messaging#
\bibitem{MSN} \verb#http://en.wikipedia.org/wiki/Windows_Live_Messenger#
\bibitem{IRC1} \verb#http://ircnet.irchelp.org/#
\bibitem{IRC2} \verb#http://rfc.net/rfc2812.html#
\bibitem{IRN} \verb#http://www.ircnet.org/#
\bibitem{NTP} \verb#http://www.ntp.org/#
\bibitem{UNR} \verb#http://www.unrealircd.com/#
\bibitem{UND} \verb#http://www.undermind.net/#
\end{thebibliography}
\end{document}